\documentclass[article,nofootinbib]{revtex4-2}

\usepackage{amsmath}
\usepackage{graphicx}
\usepackage{epsf}
\usepackage{psfrag}
\usepackage{epsfig}
\usepackage{graphics}
\usepackage{amsfonts}
\usepackage{epstopdf}
\usepackage{slashed}
\usepackage{footnote}
\usepackage[utf8]{inputenc}
\usepackage[T1]{fontenc}
\usepackage[makeroom]{cancel}
\usepackage{tikz}
\def\checkmark{\tikz\fill[scale=0.4](0,.35) -- (.25,0) -- (1,.7) -- (.25,.15) -- cycle;} 

\begin{document}
	
	\newcommand{\be}{\begin{equation}}
		\newcommand{\ee}{\end{equation}}
	\newcommand{\bq}{\begin{eqnarray}}
		\newcommand{\eq}{\end{eqnarray}}
	\newcommand{\ba}{\begin{align}}
		\newcommand{\ea}{\end{align}}
	
	\newcommand{\Dslash}{\hbox{$\partial\!\!\!{\slash}$}}
	\newcommand{\qslash}{\hbox{$q\!\!\!{\slash}$}}
	\newcommand{\pslash}{\hbox{$p\!\!\!{\slash}$}}
	\newcommand{\bslash}{\hbox{$b\!\!\!{\slash}$}}
	\newcommand{\kslash}{\hbox{$k\!\!\!{\slash}$}}
	\newcommand{\kbruto}{\hbox{$k \!\!\!{\slash}$}}
	\newcommand{\pbruto}{\hbox{$p \!\!\!{\slash}$}}
	\newcommand{\qbruto}{\hbox{$q \!\!\!{\slash}$}}
	\newcommand{\lbruto}{\hbox{$l \!\!\!{\slash}$}}
	\newcommand{\bbruto}{\hbox{$b \!\!\!{\slash}$}}
	\newcommand{\parbruto}{\hbox{$\partial \!\!\!{\slash}$}}
	\newcommand{\Abruto}{\hbox{$A \!\!\!{\slash}$}}
	\newcommand{\bbbruto}{\hbox{$b_1 \!\!\!{\slash}$}}
	\newcommand{\bbbbruto}{\hbox{$b_2 \!\!\!{\slash}$}}

	\title{Gauge symmetry and radiatively induced terms in dimension-5 non-minimal Lorentz-violating QED}
	
	\author{A. P. B. Scarpelli$^{(a)}$, A. R. Vieira$^{(b)}$}
	
	\affiliation{(a) Centro Federal de Educa\c{c}\~ao Tecnol\'ogica, Belo Horizonte -MG, Brazil\\
		(b) Universidade Federal do Tri\^angulo Mineiro, Instituto de Ci\^encias Agr\'arias, Exatas e Biol\'ogicas - ICAEBI,38280-000, Iturama- MG, Brasil}
	\date{\today}
	
	\begin{abstract}
		In this work, we derive the conditions that assure gauge invariance of a non-minimal dimension-5 Lorentz-violating QED. The two and three point functions at one-loop are computed. The gauge Ward identities are checked and the conditions to assure gauge symmetry of this non-minimal framework is found to be the same of the usual QED. Induced terms are also investigated and it is shown that
		the non-minimal Lorentz-violating $a^{(5)}_F$-term  of the fermion sector can induce radiatively a non-minimal Lorentz-violating term in the photon sector.
	\end{abstract}
	
	
	\maketitle
	
\section{Introduction}
\label{s1}

    Lorentz and CPT Symmetries are main ingredients to build quantum field theories and are usually taken for granted. Nevertheless, in string theory, it was realized long ago that these symmetries can be spontaneously broken at Planck scale and as a result Lorentz and CPT-violating terms appear at low energies \cite{Samuel}. All possible Lorentz and CPT- violating operators at these low energies make up the Standard Model Extension (SME) \cite{{Alan1},{Alan0}}. Even if space-time symmetries are in fact exact in nature, the issue is on what precision one can say they are. In this way, the SME is also a framework to test how good are these	symmetries besides being a way for searching for Lorentz and CPT violation.
    
	Precise tests of space-time symmetries generally involve classical Lagrangians or tree level Feynman diagrams of the SME.	However, there are many theoretical studies concerning this framework beyond tree level. The SME was shown to be renormalizable at one-loop for both the electroweak \cite{{Alan2},{Colladay3}} and the strong sectors \cite{{Colladay},{Colladay4}}. There are also 
several investigations concerning the radiatively induced finite quantum corrections. The most known example is the dimension-3 
operator associated with the CPT and Lorentz-violating coefficient $b_{\mu}$ from the fermion sector, which radiatively induces the 
Chern-Simons-like term of the photon sector. Furthermore, breaking Lorentz and CPT symmetries at the classical level does not mean that all the possible violating structures at this level also appear at the quantum one \cite{IJMPA}. Thus, it is worth to investigate
what symmetries can be broken at the quantum level if the Lorentz and CPT ones are broken at the classical one.

	At the same time, from a technical point of view, the regularization procedure is required beyond tree level to treat ultraviolet 
and/or infrared divergences. As it is well known, the treatment of these infinities can spuriously break symmetries of the model. For 
instance, lattice regularization is developed for treating QCD non-perturbatively and discretization of the space-time points breaks 
Lorentz symmetry among others \cite{Costa}. The renormalization process is then more laborious because the introduction of restoring 
counter-terms is needed. In the study of anomalies, this issue is more subtle because we want to know if the symmetry is indeed 
physical or if it was broken by the regularization scheme. Some anomalies are measurable, like the chiral anomaly or the scale anomaly 
in QED and this experimental fact solves the issue. The former is related with the pion decay into two photons \cite{{Jackiw}, 
{Adler}, {Bardeen}} and the later with the hadronic $R$ ratio \cite{Ellis}. On the other hand, there are not yet observables related 
with Super-symmetric anomalies or anomalies in the Standard Model Extension. Thus, the question if a symmetry is in 
fact broken at the quantum level can only be answered by using a regularization scheme that does not break spuriously any symmetry of the model. The issue of the chiral anomaly in a Lorentz-violating context is particularly discussed in \cite{{AoPCALV},{AntAn}}.
	
    Gauge symmetry is one of the ways of introducing particle interactions. Maintaining this symmetry beyond tree level assures 
renormalizability of the theory and a massless photon. Therefore, the search for a gauge invariant regularization method is desired mainly for gauge field theories, like dimensional regularization. However, as new theories arose, other proposal of regularization schemes did as well, for dealing with issues that conventional regularizations could not deal. In this work, we show that there is no gauge anomaly at one-loop for a non-minimal dimension-5 Lorentz-violating version of QED. If Bose symmetry is not required, there would be a gauge anomaly due to the three point function diagrams and the induced terms could be used to put stringent constraints in a set of non-minimal dimension-5 Lorentz and CPT-violating coefficients since gauge symmetry breaking is not observed. The paper is divided as follows: in section \ref{sIR}, we present an overview on the implicit regularization scheme. In section \ref{sGI}, we discuss the conditions for gauge invariance in the usual QED. In section \ref{sDGI}, we compute the gauge Ward identities of the two and the three point Green functions of a non-minimal Lorentz-violating QED considering the $a^{(5)}_F$-term, using both dimension and implicit regularizations. In section \ref{Ss4}, we compute these same identities considering the non-minimal $b^{(5)}_F$-term and we present conclusions in section \ref{sCP}.

	\section{Overview of Implicit Regularization}
	\label{sIR}
	
	Besides using dimensional regularization to compute divergent integrals in amplitudes of the next sections, we also apply the implicit regularization scheme \cite{IR}. The former is probably the most known and popular regularization scheme, and it does not need
	an introduction. The latter, on the other hand, although used in a wide variety of problems, is not known in textbooks.
	
	A particularly interesting application of Implicit regularization occurs when the theories include dimension specific objects, like $\gamma^5$ matrices and Levi-Civita 
	symbols. Also, since it is a scheme that does not break symmetries of the theory, it is usually used for the computation of anomalies. A 
	recent computation for a general momentum routing concerns gravitational anomalies in two dimensions 
	\cite{Orimar}. It was used also in other scenarios with Lorentz violation, like in the Bumblebee model \cite{Ricardo}, or chiral models \cite{Ricardo2}, that deal directly with $\gamma^5$ matrices, in which comparisons with other regularization techniques are performed \cite
	{{Ricardo},{Ricardo2},{Adriano3}}. 
	
	Let us make a brief review of the method in four dimensions. In this scheme, we assume that the integrals are regularized by an implicit regulator $\Lambda$ in order to allow algebraic  
	operations within the integrands. We then recursively use the following identity 
	\begin{equation}
		\frac{1}{(k+p)^2-m^2} = \frac{1}{k^2-m^2}
		-\frac{(p^2+2p\cdot k)}{(k^2-m^2)[(k+p)^2-m^2]},
		\label{2.1}
	\end{equation}
	to separate basic divergent integrals (BDI's) from the finite part. These BDI's are defined as follows
	\begin{equation}
		I^{\mu_1 \cdots \mu_{2n}}_{log}(m^2)\equiv \int_k \frac{k^{\mu_1}\cdots k^{\mu_{2n}}}{(k^2-m^2)^{2+n}}
	\end{equation}
	and
	\begin{equation}
		I^{\mu_1 \cdots \mu_{2n}}_{quad}(m^2)\equiv \int_k \frac{k^{\mu_1}\cdots k^{\mu_{2n}}}{(k^2-m^2)^{1+n}}.
	\end{equation}
	
	The BDI's with Lorentz indices can be judiciously combined as differences between integrals with the same superficial degree 
	of divergence, according to the equations below, which define surface terms  \footnote{The Lorentz indices between brackets stand for 
		permutations, i.e. $A^{\{\alpha_1\cdots\alpha_n\}}B^{\{\beta_1\cdots\beta_n\}}=A^{\alpha_1\cdots\alpha_{n}}B^{\beta_1\cdots\beta_n}$ 
		+ sum over permutations between the two sets of indices $\alpha_1\cdots\alpha_{n}$ and $\beta_1\cdots\beta_n$. For instance, $g^{\{\mu\nu}g^{\alpha\beta\}}=
		g^{\mu\nu}g^{\alpha\beta}+g^{\mu\alpha}g^{\nu\beta}+g^{\mu\beta}g^{\nu\alpha}$.}:
	\begin{eqnarray}
		\Upsilon^{\mu \nu}_{2w}=  g^{\mu \nu}I_{2w}(m^2)-2(2-w)I^{\mu \nu}_{2w}(m^2) 
		\equiv \upsilon_{2w}g^{\mu \nu},
		\label{dif1}\\
		\nonumber\\
		\Xi^{\mu \nu \alpha \beta}_{2w}=  g^{\{ \mu \nu} g^{ \alpha \beta \}}I_{2w}(m^2)
		-4(3-w)(2-w)I^{\mu \nu \alpha \beta }_{2w}(m^2)\equiv\nonumber\\
		\equiv  \xi_{2w}(g^{\mu \nu} g^{\alpha \beta}+g^{\mu \alpha} g^{\nu \beta}+g^{\mu \beta} g^{\nu \alpha}).
		\label{dif2}
	\end{eqnarray} 
	
	In the expressions above, $2w$ is the degree of divergence of the integrals and we adopt the notation such that indices $0$ and $2$ mean $log$ and $quad$,
	respectively. Surface terms can be conveniently written as integrals of total derivatives, as presented below:
	
	\begin{eqnarray}
		\upsilon_{2w}g^{\mu \nu}= \int_k\frac{\partial}{\partial k_{\nu}}\frac{k^{\mu}}{(k^2-m^2)^{2-w}}, \nonumber \\
		\label{ts1}
	\end{eqnarray}
	\begin{eqnarray}
		(\xi_{2w}-v_{2w})(g^{\mu \nu} g^{\alpha \beta}+g^{\mu \alpha} g^{\nu \beta}+g^{\mu \beta} g^{\nu \alpha})= \int_k\frac{\partial}{\partial 
			k_{\nu}}\frac{2(2-w)k^{\mu} k^{\alpha} k^{
				\beta}}{(k^2-m^2)^{3-w}}.
		\label{ts2}
	\end{eqnarray}
	
	We see that the surface terms in equations (\ref{dif1})-(\ref{dif2}) are undetermined because they are differences between divergent quantities. Each 
	regularization scheme gives a different value for these terms. However, as physics should not depend on the scheme applied, we leave these terms to be arbitrary until the end of the calculation and then fix them by symmetry 
	constraints or phenomenology. This approach was first proposed in \cite{JackiwFU}, where undetermined surface terms were discussed
	in several contexts of quantum corrections.
	
	Of course, the same idea can be applied for any dimension of space-time and for higher loops. Equation (\ref{2.1}) is used recursively 
	until the divergent piece is separated from the finite one. This procedure makes the finite integrals hard to compute due to the 
	number of $k$'s in the numerator. A simpler alternative to this approach is presented in \cite{Bruno}, where the Feynman 
	parametrization is applied before separating the BDI's. Also, eq. (\ref{2.1}) is not the only possible equation to be used since the 
	implicit regulator was assumed to allow the use of other identities.
	
	\subsection{An example: Gauge invariance of the vacuum polarization tensor}
	\label{sGI}
	
	Let us consider the vacuum polarization tensor of conventional Quantum Electrodynamics (QED) whose computation in implicit regularization is given by (all regularized integrals are presented in the appendix)
	\bq
	\centering
	i\Pi^{\mu\nu}(p)&=\frac{4}{3}e^2(p^2g^{\mu\nu}-p^{\mu}p^{\nu})I_{log}(m^2)-4e^2\upsilon_2g^{\mu\nu}+\frac{4}{3}e^2(p^2g^{\mu\nu}-p^{\mu}p^{\nu})\upsilon_0 \nonumber\\
	&-\frac{4}{3}e^2(p^2g^{\mu\nu}+2p^{\mu}p^{\nu})(\xi_0-2\upsilon_0)-\frac{i}{2\pi^2}e^2(p^2g^{\mu\nu}-p^{\mu}p^{\nu})(Z_1-Z_2),
	\label{eqpi}
	\eq
	in which $Z_n=\int_0^1 dx x^n\ln \left(\frac{D(x)}{m^2}\right)$ and $D(x)=m^2-p^2 x(1-x)$ and where, for didactic reasons, we have placed the divergent, the finite and surface-dependent terms separately.
	
\begin{figure}[!h]
		\centering
		\includegraphics[trim=0mm 20mm 0mm 20mm , width=0.5\textwidth]{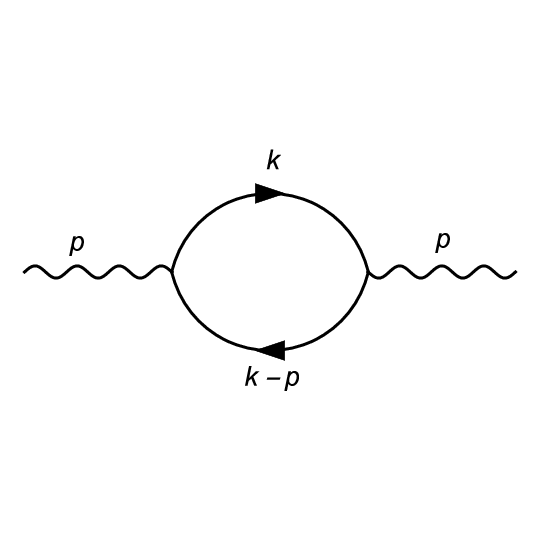}
		\caption{Vacuum polarization tensor of QED.}
		\label{fig0}
	\end{figure}

	Notice that if we require gauge invariance using the Ward identity $p_{\mu}\Pi^{\mu\nu}(p)=0$, we find that the quadratic surface term 
	$\upsilon_2$ must be zero and that the logarithmic surface terms must obey the relation $\xi_0=2\upsilon_0$. These conditions are automatically fulfilled if we set all surface terms to zero. The same takes place when one uses dimensional regularization, because the surface terms defined in section \ref{sIR} are zero in such a scheme. This is the same condition 
	obtained if we require momentum routing invariance of the Feynman diagram in figure \ref{fig0} (this feature of the loop diagram 
	would not appear in dimensional regularization, since it allows for shifts in the loop momenta). In the next sections, we are going to 
	see that these requirements for surface terms are the same for a non-minimal dimension-5 Lorentz-violating version of QED. The Ward 
	identities are checked for two and three-point functions in this Lorentz-violating framework.
	
	The diagrams with more external photon legs in usual QED do not need to be checked as the ones of the next sections. The three photon
	leg of the usual QED is zero because of Furry's theorem, and the box diagram with four photon legs is gauge invariant, since there is no term in the tree level Lagrangian to renormalize it if it was not.  
	
	\section{Gauge Invariance and Radiatively Induced terms: $a^{(5)}_F$-term }
	\label{sDGI}
	
\begin{figure}\centering
		\includegraphics[trim=0mm 30mm 0mm 30mm ,scale=0.8]{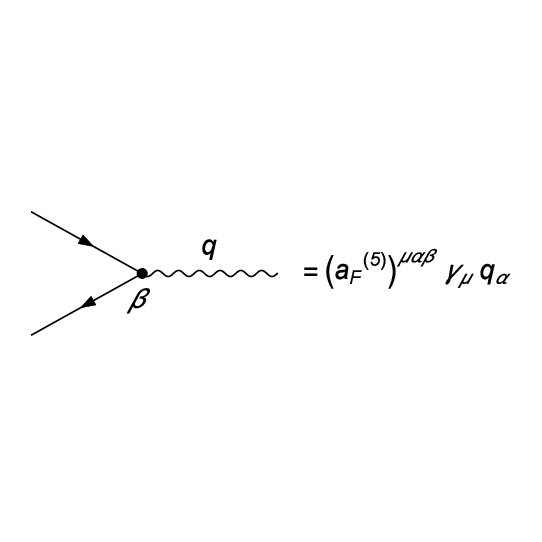}
		\caption{Feynman rule for a non-minimal Lorentz-violating interaction.}
		\label{fig1}
	\end{figure}

	We consider the dimension-5 coefficients in the Lagrangian density $\mathcal{L}^{(5)}_{\psi F}$ of Table I of reference \cite{Zonghao}. The terms with coefficients $(m_F^{(5)})^{\alpha\beta}$, $(m_{5F}^{(5)})^{\alpha\beta}$ and $(b_F^{(5)})^{\mu\alpha\beta}$ do not generate
	induced terms at first order because the number of Dirac matrices in the trace of the fermion loop is odd or there are less than
	four Dirac matrices appearing with a $\gamma_5$ matrix inside this trace. The non-minimal term $(H_F^{(5)})^{\mu\nu\alpha\beta}\overline{\psi}\sigma_{\mu\nu}F_{\alpha\beta}\psi$ was studied in \cite{Manoel}, in which the authors showed that it radiatively induces the CPT-even term $(k_F)^{\mu\nu\alpha\beta}F_{\alpha\beta}F_{\mu\nu}$ of the minimal SME photon sector. The other non-trivial radiatively induced term comes from the CPT and Lorentz-violating non-minimal term
	$-\frac{1}{2}(a^{(5)}_F)^{\mu\alpha\beta}\overline{\psi}\gamma_{\mu}F_{\alpha\beta}\psi$. It can be rewritten as $-(a^{(5)}_F)^{\mu\alpha\beta}\overline{\psi}\gamma_{\mu}\partial_{\alpha}A_{\beta}\psi$ due to the antisymmetry of the two last indices 
	and this leads to the Feynman rule presented in Figure \ref{fig1}. It is also important to notice that it is easy to check with
	the modified version of the Dirac equation that this term does not break gauge symmetry at the classical level,  {\it i. e.} 
	$\partial_{\mu}j^{\mu}=0$, where $j^{\mu}=\bar{\psi}\gamma^{\mu}\psi$.
	
	We next perform perturbative calculations in a modified Lorentz-violating QED model which includes this term.
	
	\subsection{Two-point function}
	
\begin{figure}\centering
		\includegraphics[trim=0mm 30mm 0mm 30mm ,scale=0.8]{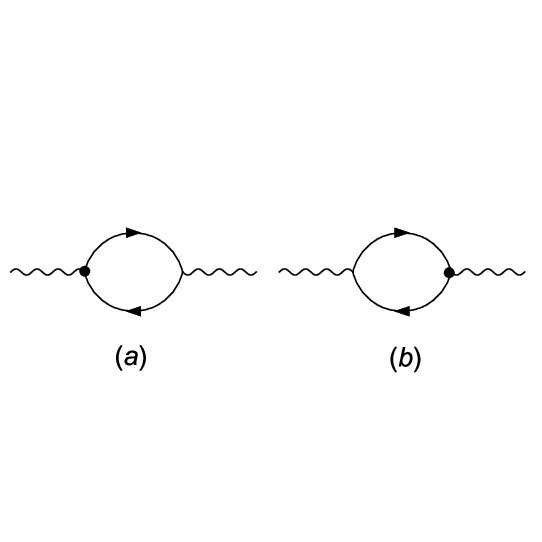}
		\caption{One-loop diagrams contributing for the radiatively induction of a non-minimal Lorentz-violating term.}
		\label{fig2}
	\end{figure}

	The diagrams depicted in Figure \ref{fig2} give rise to a radiatively induced non-minimal LV term in the photon sector. In order to 
	see this, we compute the diagrams of this figure. Their corresponding amplitudes can be written as 
	
	\begin{align}
		&\Pi^{\alpha\beta}_{(a)}(p)=-\int_k Tr\left[(a^{(5)}_F)^{\mu\lambda\beta}\gamma_{\mu}p_{\lambda}\frac{i}{\slashed{k}-\slashed{p}-m}
		(-ie \gamma^{\alpha})\frac{i}{\slashed{k}-m} \right] 
	\end{align}
	and
	\begin{align}
		&\Pi^{\alpha\beta}_{(b)}(p)=-\int_k Tr\left[(-ie \gamma^{\beta})\frac{i}{\slashed{k}-\slashed{p}-m}(a^{(5)}_F)^{\mu\lambda\alpha}\gamma_{\mu}p_{\lambda}\frac{i}{\slashed{k}-m} \right],
		\label{eq2.1}
	\end{align}
	where $\int_k$ stands for $\int \frac{d^4k}{(2\pi)^4}$. Let us then apply Implicit Regularization, presented in section \ref{sIR}, in the computation of the diagrams of the two-point function. After taking the traces and regularizing (all integrals are presented in appendix), we find the following result:
	\begin{align}
		&\Pi^{\alpha\beta}(p)=\Pi^{\alpha\beta}_{(a)}(p)+\Pi^{\alpha\beta}_{(b)}(p)=\frac{4}{3}i((a_F^{(5)})^{p p\beta}p^{\alpha}-
		(a_F^{(5)})^{\alpha p \beta}p^{2})\left( I_{log}(m^2)- b Z_0-\frac{i}{48\pi^2}\right)+ \nonumber\\
		&+\frac{8i b }{3}m^2 Z_0\left( -(a_F^{(5)})^{\alpha p\beta}+\frac{p^{\alpha}}{p^2}(a_F^{(5)})^{p p\beta}\right) +4( (a_F^{(5)})^{p p\beta}p^{\alpha}+(a_F^{(5)})^{\alpha p \beta}p^{2})\upsilon_0-\nonumber\\
		&-\frac{4}{3}(-2(a_F^{(5)})^{p p \beta}p^{\alpha}+(a_F^{(5)})^{\alpha p\beta}p^{2})\xi_0+(\alpha \leftrightarrow \beta),
		\label{eq2.2}
	\end{align}
	where $(a^{(5)}_F)^{\alpha p\beta}\equiv (a^{(5)}_F)^{\alpha \mu\beta}p_{\mu}$ and $b=\frac{i}{(4\pi)^2}$. We can easily check gauge invariance by computing the Ward identity $p_{\alpha}\Pi^{\alpha\beta}(p)=0$. Doing this, we find out that the surface terms break gauge symmetry, as expected:
	\begin{align}
		p_{\alpha}\Pi^{\alpha\beta}(p)=4(a_F^{(5)})^{p p \beta}p^2(-\xi_0+2\upsilon_0)-4\upsilon_2 (a_F^{(5)})^{p p \beta}
		\label{eq2.3}
	\end{align}
	
	We see above that, if all surface terms are null, gauge symmetry is automatically fulfilled. However, the relations $\xi_0=2\upsilon_0$ and $\upsilon_2=0$ are sufficient. It is interesting to notice that 
	these are the same conditions for gauge invariance presented in section \ref{sGI} and found in other frameworks like in the minimal QED extension \cite{IJMPA}.
	
	Alternatively, one can apply dimensional regularization in this calculation. The result is the same as in Implicit Regularization as long as the surface terms of eq. (\ref{eq2.2}) are set to zero and we take
	\begin{equation}
		I_{log}(m^2)= b\left(\frac{1}{\epsilon}-\frac 12 \ln\left(-\frac{m^2}{4\pi\mu^2}\right) -\frac{\gamma}{2}\right),
	\end{equation}
	in which $\mu$ is a mass scale introduced to keep the number of dimensions of the integrals after changing it to $d$ dimensions. This result is expected, since surface terms are zero in dimensional regularization.
	
	In the following discussion, we are interested in the form of the induced term, so we are going to assume for the moment that the surface terms are equal to zero. The non-minimal term of the fermion sector radiatively induces a term in the photon sector which is also non-minimal. In order to see this, let us consider the effective action $S=\frac{1}{2}\int d^4 x A_{\alpha}\Pi^{\alpha\beta}A_{\beta}$ and take the massless limit of eq. (\ref{eq2.2}). It leads to 
	
	\begin{align}
		& S=\frac{4i}{3}\int d^4 x\left\{ (a_F^{(5)})^{\alpha \mu \beta}A_{\alpha}\partial^2\partial_{\mu}A_{\beta} -(a_F^{(5)})^{\mu \nu \beta}A_{\alpha}\partial^{\alpha}\partial_{\mu}\partial_{\nu}A_{\beta}\right\}\left( I_{log}(\lambda^2)-b \ln \left(\frac{-p^2}{\lambda^2} 
		\right) -\frac{b}{3}\right) =\nonumber\\
		& =\frac{2}{3}\int d^4 x \ i(a^{(5)}_F)^{\alpha \mu \beta}F_{\nu\alpha}\partial^{\nu}F_{\mu\beta}\left( I_{log}(\lambda^2)-b \ln \left(\frac{-p^2}{\lambda^2} \right) -\frac{b}{3}\right),
		\label{eq2.4}
	\end{align}
	where we dropped out terms that are total derivatives. The induced term has the form $(a^{(5)}_F)^{\alpha \mu \beta}F_{\nu\alpha}\partial^{\nu}F_{\mu\beta}$, that is a particular case of
	\begin{equation}
		{\cal L}_A^{(5)}= -\frac 14 k^{(5) \alpha\kappa\lambda\mu\nu}F_{\kappa\lambda}\partial_\alpha F_{\mu\nu},
	\end{equation}
	present in Table III of reference \cite{Zonghao}, in which $k^{(5) \alpha\kappa\lambda\mu\nu}$ is proportional to
	\begin{equation}
		\eta^{\alpha\lambda} (a^{(5)}_F)^{\kappa\mu\nu} - \eta^{\alpha\kappa} (a^{(5)}_F)^{\lambda\mu\nu} + \eta^{\alpha\mu} (a^{(5)}_F)^{\nu\kappa\lambda} - \eta^{\alpha\nu} (a^{(5)}_F)^{\mu\kappa\lambda}.
	\end{equation}
	
	Eq. (\ref{eq2.4}) shows us the presence of a divergent part in the coefficient of the induced term, which is an important point to be analyzed. This indicates that the original classical action must contain such a term. In other words, the
	inclusion of the dimension-5 term $-\frac{1}{2}(a^{(5)}_F)^{\mu\alpha\beta}\overline{\psi}\gamma_{\mu}F_{\alpha\beta}\psi$ in a modified QED requires the presence of this induced term from the beginning. However, we must take into account that our model is non-renormalizable. We have carried out a one-loop calculation and, at this order in the perturbative expansion, it has been shown that a new term which violates Lorentz and CPT symmetries should be included in the classical action. If we go
	beyond the one-loop order, certainly new other terms will have to be considered. The non-renormalizability
	of the model tells us that there is not a finite number of counter-terms that will be sufficient to renormalize
	the theory. So, if we would like to deal with this effective model, we will have to stop at one-loop
	order. For this, it is necessary to find a cutoff energy $\Lambda$. This discussion is carried out in \cite{scarp} and \cite{scarp2}.			
	Finally, it is easy to check that this term leads to the usual charge conservation at the classical level. The modified Maxwell 
	equations for an additional $C (a^{(5)}_F)^{\alpha \mu \beta}F_{\nu\alpha}\partial^{\nu}F_{\mu\beta}$ are 
	$\partial_{\lambda}F^{\lambda\zeta}=J^{\zeta}+C((a^{(5)}_F)^{\zeta \mu \beta}\partial^2F_{\mu\beta}-(a^{(5)}_F)^{\lambda \mu 
		\beta}\partial_{\lambda}\partial^{\zeta}F_{\mu\beta})$, where we can easily see that 
	$\partial_{\zeta}\partial_{\lambda}F^{\lambda\zeta}=\partial_{\zeta}J^{\zeta}=0$.
	
	\subsection{Three-point function}
	
\begin{figure}\centering
		\includegraphics[trim=0mm 70mm 0mm 70mm ,scale=0.8]{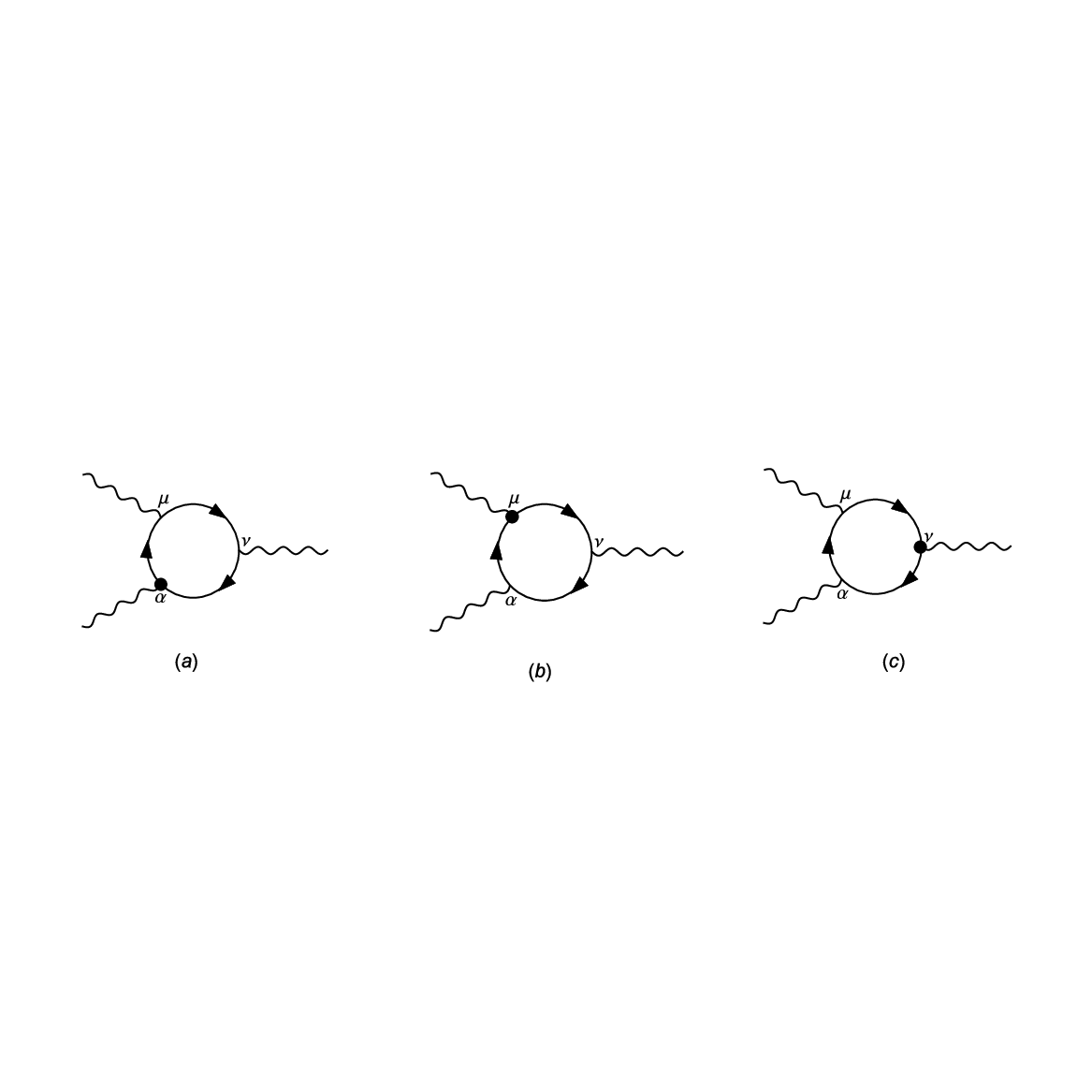}
		\caption{One-loop diagrams with three photon legs.}
		\label{fig3}
	\end{figure}

	The next step would be to check gauge symmetry for
	diagrams with more photon external legs. Let us consider the diagrams of the three-point function presented in Figure \ref{fig3}. Using the Feynman rules set out above, the corresponding amplitudes 
	can be written as below:
	\begin{align}
		&T^{\mu\nu\alpha}_{(a)}(p,q)=-\int_k Tr\left[(-ie\gamma^{\mu})\frac{i}{\slashed{k}-m}(-ie\gamma^{\nu})\frac{i}{\slashed{k}+\slashed{q}-m} (a_F^{(5)})^{\lambda\zeta\alpha}(p+q)_{\zeta}\gamma_{\lambda}\frac{i}{\slashed{k}-\slashed{p}-m}\right], \nonumber\\
		&T^{\mu\nu\alpha}_{(b)}(p,q)=-\int_k Tr\left[(a_F^{(5)})^{\lambda\zeta\mu}p_{\zeta}\gamma_{\lambda}\frac{i}{\slashed{k}-m}(-ie\gamma^{\nu})\frac{i}{\slashed{k}+\slashed{q}-m}(-ie\gamma^{\alpha}) \frac{i}{\slashed{k}-\slashed{p}-m}\right] \nonumber\\
		\textrm{and}\\
		&T^{\mu\nu\alpha}_{(c)}(p,q)=-\int_k Tr\left[(-ie\gamma^{\mu})\frac{i}{\slashed{k}-m}(a_F^{(5)})^{\lambda\zeta\nu}q_{\zeta}\gamma_{\lambda}\frac{i}{\slashed{k}+\slashed{q}-m} (-ie\gamma^{\alpha})\frac{i}{\slashed{k}-\slashed{p}-m}\right].
		\label{eq2.5}
	\end{align}
	Besides, in virtue of Bose symmetry, the amplitudes referring to the crossed diagrams have to be added. We then have
	\begin{equation}
		T^{\mu\nu\alpha}(p,q) = T^{\mu\nu\alpha}_{(a)}(p,q) + T^{\mu\nu\alpha}_{(b)}(p,q) + T^{\mu\nu\alpha}_{(c)}(p,q) + \mbox{crossed terms}
	\end{equation}
	
	It is easy to check gauge invariance for diagram $(a)$. We get $(p+q)_{\alpha}T^{\mu\nu\alpha}_{(a)}(p,q)=0$ due to the antisymmetry 
	property of the two last indices of the tensor $(a_F^{(5)})^{\lambda\zeta\alpha}$. For the next two diagrams, we use the identity $\slashed{q}+\slashed{p}=\slashed{k}+\slashed{q}-m
	-(\slashed{k}-\slashed{p}-m)$ after the contraction in order to split the amplitude into two pieces. After this manipulation, we find:
	\begin{align}
		&(p+q)_{\alpha}T^{\mu\nu\alpha}_{(b)}(p,q)=e^2(a_F^{(5)})^{\lambda\zeta\mu}p_{\zeta}\Big\{ -\int_k \frac{Tr[\gamma_{\lambda}(\slashed{k}+m)\gamma^{\nu}(\slashed{k}-\slashed{p}+m)]}{(k^2-m^2)[(k-p)^2-m^2]}+ \nonumber\\
		&+\int_k \frac{Tr[\gamma_{\lambda}(\slashed{k}+m)\gamma^{\nu}(\slashed{k}+\slashed{q}+m)]}{(k^2-m^2)[(k+q)^2-m^2]}\Big\}\nonumber\\
		\textrm{and}\\
		&(p+q)_{\alpha}T^{\mu\nu\alpha}_{(c)}(p,q)=e^2(a_F^{(5)})^{\lambda\zeta\nu}q_{\zeta}\Big\{ -\int_k \frac{Tr[\gamma^{\mu}(\slashed{k}+m)\gamma_{\lambda}(\slashed{k}-\slashed{p}+m)]}{(k^2-m^2)[(k-p)^2-m^2]}+ \nonumber\\
		&+\int_k \frac{Tr[\gamma^{\mu}(\slashed{k}+m)\gamma_{\lambda}(\slashed{k}+\slashed{q}+m)]}{(k^2-m^2)[(k+q)^2-m^2]}\Big\}.
		\label{eq2.6}
	\end{align}

	The computation of equations (\ref{eq2.6}) with implicit regularization reveals that the crossed terms are necessary in order to avoid a gauge anomaly for the 
	three-point diagram, even if all the surface terms are zero. The result is given by
	\begin{align}
		&(p+q)_{\alpha}\left(T^{\mu\nu\alpha}_{(a)} + T^{\mu\nu\alpha}_{(b)} + T^{\mu\nu\alpha}_{(c)}\right)= \nonumber \\
		&=\frac{4}{3}\left(I_{log}(\lambda^2 )-b \ln \left(\frac{-p^2}{
			\lambda^2}\right)-\frac{b}{3} \right)\left( p^{\nu}(a^{(5)}_F)^{pp\mu}-p^2(a^{(5)}_F)^{\nu p\mu}+
		p^{\mu}(a^{(5)}_F)^{pq\nu}-p^2(a^{(5)}_F)^{\mu p\nu}\right)+\nonumber\\
		&+\frac{4}{3}\xi_0(2p^{\nu}(a^{(5)}_F)^{pp\mu}+p^2(a^{(5)}_F)^{\nu p\mu}+2p^{\mu}(a^{(5)}_F)^{pq\nu}+p^2(a^{(5)}_F)^{\mu p\nu})
		-4\upsilon_0( p^{\nu}(a^{(5)}_F)^{pp\mu}+\nonumber\\
		&+p^2(a^{(5)}_F)^{\nu p\mu}+p^{\mu}(a^{(5)}_F)^{pq\nu}+p^2(a^{(5)}_F)^{\mu p\nu}).
		\label{eq2.7}
	\end{align}
	When the crossed terms are added, the gauge symmetry, in this case, is fulfilled in a way that is independent of the surface terms.

	\section{Gauge Invariance: non-minimal dimension-5 $b^{(5)}_F$-term}
	\label{Ss4}
	
	The term of coefficient $(b_F^{(5)})^{\mu\alpha\beta}$ does not generate induced terms at first order due to the anti-symmetry properties of the Levi-Civita symbol. It is also easy to check the gauge 
	Ward identities for the 2-point function. The diagrams are the same as Figure \ref{fig2} except for the different vertex and the amplitudes are given by:
	\begin{align}
		&A^{\alpha\beta}_{(a)}(p)=-\int_k Tr\left[(b^{(5)}_F)^{\mu\lambda\beta}\gamma_{5}\gamma_{\mu}p_{\lambda}\frac{i}{\slashed{k}-\slashed{p}-m}
		(-ie \gamma^{\alpha})\frac{i}{\slashed{k}-m} \right] \nonumber\\
		\textrm{and}\\
		&A^{\alpha\beta}_{(b)}(p)=-\int_k Tr\left[(-ie \gamma^{\beta})\frac{i}{\slashed{k}-\slashed{p}-m}(b^{(5)}_F)^{\mu\lambda\alpha}\gamma_{5}\gamma_{\mu}p_{\lambda}\frac{i}{\slashed{k}-m} \right].
		\label{eq3.1}
	\end{align}
	
	When computing $p_{\alpha}A^{\alpha\beta}(p)=p_{\alpha}A^{\alpha\beta}_{(a)}(p)+p_{\alpha}A^{\alpha\beta}_{(b)}(p)$,
	the second term vanishes because of the antisymmetry of the last two indices of the tensor $b^{(5)}_F$. For the
	first term, we can use $\slashed{p}=\slashed{k}-m-(\slashed{k}-\slashed{p}-m)$ to split the integral into two pieces. Each one of these pieces is zero, because there remain two Dirac matrices with one $\gamma_5$ inside the trace.
	
	It is also important to check with the use of the modified version of the Dirac equation that this non-minimal term does not break gauge or chiral symmetries at the classical level, {\it i. e.} $\partial_{\mu}j^{\mu}=0$ and $\partial_{\mu}j^{\mu}_{5}=0$, where
	$j^{\mu}=\bar{\psi}\gamma^{\mu}\psi$ and $j^{\mu}_5=\bar{\psi}\gamma^{\mu}\gamma_5\psi$. We could expect a chiral anomaly due to 
	the presence of a $\gamma_5$ matrix in one of the vertices of the 3-point function diagram. However, this is not the case as it is shown below and there is no issue with $\gamma_5$ in the regularization applied.
	
	The diagrams with three photon legs are the same ones depicted in Figure \ref{fig3}, the only difference being the Lorentz-violating vertex. The amplitude corresponding to these diagrams can be written as:
	\begin{align}
		&M^{\mu\nu\alpha}_{(a)}(p,q)=-\int_k Tr\left[(-ie\gamma^{\mu})\frac{i}{\slashed{k}-m}(-ie\gamma^{\nu})\frac{i}{\slashed{k}+\slashed{q}-m} (b_F^{(5)})^{\lambda\zeta\alpha}(p+q)_{\zeta}\gamma_5\gamma_{\lambda}\frac{i}{\slashed{k}-\slashed{p}-m}\right], \nonumber\\
		&M^{\mu\nu\alpha}_{(b)}(p,q)=-\int_k Tr\left[(b_F^{(5)})^{\lambda\zeta\mu}p_{\zeta}\gamma_5\gamma_{\lambda}\frac{i}{\slashed{k}-m}(-ie\gamma^{\nu})\frac{i}{\slashed{k}+\slashed{q}-m}(-ie\gamma^{\alpha}) \frac{i}{\slashed{k}-\slashed{p}-m}\right] \nonumber\\
		\textrm{and}\\
		&M^{\mu\nu\alpha}_{(c)}(p,q)=-\int_k Tr\left[(-ie\gamma^{\mu})\frac{i}{\slashed{k}-m}(b_F^{(5)})^{\lambda\zeta\nu}q_{\zeta}\gamma_5\gamma_{\lambda}\frac{i}{\slashed{k}+\slashed{q}-m} (-ie\gamma^{\alpha})\frac{i}{\slashed{k}-\slashed{p}-m}\right].
		\label{eq3.2}
	\end{align}
	
	As before, we can check the gauge Ward identities by computing $p_{\mu}M^{\mu\nu\alpha}$, for instance. Besides the
	amplitudes of eq. (\ref{eq3.2}), there are also the crossed diagrams, obtained by changing
	$\mu\leftrightarrow \nu$ and $p\leftrightarrow q$.
	
	The computation of the gauge Ward identity is a little involved and it needs the use of relations between $\xi_{nm}(p,q)$ integrals
	presented in the appendix. The idea is to reduce integrals $\xi_{20}(p,q)$, $\xi_{02}(p,q)$ and $\xi_{11}(p,q)$ into integrals 
	$\xi_{10}(p,q)$ and $\xi_{01}(p,q)$ by using eqs. (\ref{f1})-(\ref{f2}) and (\ref{f5})-(\ref{f6}). Then the remaining integrals 
	$\xi_{10}(p,q)$ and $\xi_{01}(p,q)$ are reduced into integrals $Z_n(p^2,m^2)$ and $\xi_{00}(p,q)$ with the use of eqs. (\ref{f3})-(\ref
	{f4}). Nevertheless, none of these finite integrals remain in the final result that depends only on the
	logarithmic surface term:
	\begin{equation}
		p_{\mu}T^{\mu\nu\alpha}=-\frac{1}{4\pi^2}\upsilon_0\left( (b^{(5)}_F)_{\zeta q}^{\ \ \ \nu} \epsilon^{\alpha  \zeta  p q}+2(b^{(5)}_F)_{\zeta p}^{\ \ \ \alpha}\epsilon^{\zeta  \nu  p q}+2 (b^{(5)}_F)_{\zeta q}^{\ \ \ \alpha} \epsilon ^{\zeta  \nu  p q}\right),
		\label{eq3.3}
	\end{equation}
	where $(b^{(5)}_F)_{\zeta p \alpha}\equiv(b^{(5)}_F)_{\zeta \lambda \alpha}p^{\lambda}$.
	
	In this case, we need to choose the surface term $\upsilon_0$ as zero to avoid a gauge anomaly in the three point function. This 
	condition is according to the previous ones concerning the $a^{(5)}_F$-term and the usual QED, although in this situation the
	logarithmic surface term $\upsilon_0$ appears alone.
	
	We summarized all the results of the computed gauge Ward identities in table \ref{tab1}.
	
	\begin{table}[!h]
		\centering
		\small
		\begin{tabular}{|l|l|l|l|}
			\hline
			Non-minimal term& Ward identity& Dimensional Reg. & Implicit Reg.   \\
			\hline
			Usual QED& 2-point function &  \checkmark & $\upsilon_2=0$ and $\xi_0=2\upsilon_0$ \\
			& 3-point function & Null by Furry's theorem  & Null by Furry's theorem \\
			\hline
			$a^{(5)}_F$-term& 2-point function &  \checkmark & $\upsilon_2=0$ and $\xi_0=2\upsilon_0$ \\
			& 3-point function & \checkmark  &\checkmark \\
			\hline
			$b^{(5)}_F$-term& 2-point function & \checkmark & \checkmark \\
			& 3-point function & cannot be applied &$\upsilon_0=0$ \\
			\hline
		\end{tabular}
		\caption{Summary of the conditions for gauge invariance in loop diagrams of a non-minimal dimension-5 Lorentz-violating QED.}
		\label{tab1}
	\end{table}

	\section{Conclusions}
	\label{sCP}
	
	In this work, we discussed the possibility of a gauge anomaly in a non-minimal dimension-5 Lorentz-violating framework. We showed that there is no gauge anomaly if surface terms are null or if there is a relation between logarithmic surface terms. Previous results 
in literature showed that no gauge anomalies are expected in minimal versions of the SME as well, like in the minimal Lorentz-violating QED \cite{{MRIPRD}, {Tiago}}. The same happens to be true for non-minimal versions of the SME. In particular, we explicitly showed that the non-minimal dimension-5 version of QED is gauge invariant beyond tree level.

	The gauge invariance of this non-minimal Lorentz-violating framework is expected since it was shown that the $H^{(5)}_F$-term can
induce the $k_F$ gauge invariant term in the photon sector as studied in \cite{Manoel}. Furthermore, it was shown here that the non-minimal $(a^{(5)}_F)$-term induces a non-minimal dimension-5 gauge invariant term in the photon sector. The induction of terms due to diagrams with more external legs is more involved but the corresponding Ward identities of them were checked. As a prospect, it would be interesting to investigate further loop diagrams from the non-minimal SME with higher dimension terms.
	
	\section*{Appendix}
	\label{a1}
	
	Integrals computed with implicit regularization:
	
	\begin{align}
		\centering
		&H^{\mu}_1=\int_k \frac{k^{\mu}}{(k^2-m^2)^2[(k-p)^2-m^2]}=-b p^{\mu}\iota_1, \nonumber\\
		&H^{\mu\nu}_2=\int_k \frac{k^{\mu}k^{\nu}}{(k^2-m^2)^2[(k-p)^2-m^2]}=\frac{1}{4}g^{\mu\nu}(I_{log}(m^2)-\upsilon_0-b Z_0)-b p^{\mu}p^{\nu}\iota_2, \nonumber\\
		&H^{\mu\nu\alpha}_3=\int_k \frac{k^{\mu}k^{\nu}k^{\alpha}}{(k^2-m^2)^2[(k-p)^2-m^2]}=\frac{1}{12}p^{\{ \alpha}g^{\mu\nu \}}(I_{log}(m^2)-\xi_0-6b Z_1+6b Z_2)-b p^{\mu}p^{\nu}p^{\alpha}\iota_3 \nonumber\\
		&A^{\mu}_1=\int_k \frac{k^{\mu}}{(k^2-m^2)[(k-p)^2-m^2]}=-p^2H_1^{\mu}+2p_{\lambda}H_2^{\mu\lambda} \nonumber\\
		&A^{\mu\nu}_2=\int_k \frac{k^{\mu}k^{\nu}}{(k^2-m^2)[(k-p)^2-m^2]}=\frac{1}{2}g^{\mu\nu}(I_{quad}(m^2)-\upsilon_2 )-p^2H_2^{\mu\nu}+2p_{\lambda}H_3^{\mu\nu\lambda} \nonumber\\
		&A_{0k}=\int_k \frac{1}{(k^2-m^2)[(k-p)^2-m^2]}=I_{log}(m^2)+b p^2(\iota_0 -2\iota_1) \nonumber\\
		&A_{2k}=\int_k \frac{1}{[(k-p)^2-m^2]}=I_{quad}(m^2)-p^2\upsilon_0
	\end{align}
	where $\lambda$ is the renormalization group scale, $b \equiv \frac{i}{(4\pi)^2}$, $Z_n=\int_0^1 dx x^n\ln \left(\frac{D(x)}{m^2}\right)$, $\iota_n=\int_0^1 dx \frac{x^n(1-x)}{D(x)}$ and $D(x)=m^2-p^2 x(1-x)$.
	
	\subsection*{Integrals of the finite part of the diagrams in section \ref{Ss4}}
	
	The functions $\xi_{nm}(p,q)$ are defined as
	\be
	\xi_{nm}(p,q)=\int^1_0 dz\int^{1-z}_0 dy \frac{z^n y^m}{Q(y,z)},\\
	\ee
	with
	\be
	Q(y,z)=[p^2 y(1-y)+q^2 z(1-z)+2(p\cdot q)yz-m^2]
	\ee
	and those functions have the property $\xi_{nm}(p,q)=\xi_{mn}(q,p)$.
	
	The $\xi_{nm}$ functions obey the following relations:
	\bq
	&q^2 \xi_{11}(p,q)-(p\cdot q)\xi_{02}(p,q)=\frac{1}{2}\left[ -\frac{1}{2}Z_0((p+q)^2,m^2)+\frac{1}{2}Z_0(p^2,m^2)+q^2 \xi_{01}(p,q)
	\right],\label{f1}\\
	&p^2 \xi_{11}(p,q)-(p\cdot q)\xi_{20}(p,q)=\frac{1}{2}\left[ -\frac{1}{2}Z_0((p+q)^2,m^2)+\frac{1}{2}Z_0(q^2,m^2)+p^2 \xi_{10}(p,q)
	\right],\label{f2}\\
	&q^2 \xi_{10}(p,q)-(p\cdot q)\xi_{01}(p,q)=\frac{1}{2}[ -Z_0((p+q)^2,m^2)+Z_0(p^2,m^2)+q^2 \xi_{00}(p,q)],\label{f3}\\
	&p^2 \xi_{01}(p,q)-(p\cdot q)\xi_{10}(p,q)=\frac{1}{2}[ -Z_0((p+q)^2,m^2)+Z_0(q^2,m^2)+p^2 \xi_{00}(p,q)],\label{f4}\\
	&q^2 \xi_{20}(p,q)-(p\cdot q)\xi_{11}(p,q)=\frac{1}{2}\left[-\left(\frac{1}{2}+m^2\xi_{00}(p,q)\right)+\frac{1}{2}p^2\xi_{01}(p,q)+\frac
	{3}{2}q^2\xi_{10}(p,q)\right],\label{f5}\\
	&p^2 \xi_{02}(p,q)-(p\cdot q)\xi_{11}(p,q)=\frac{1}{2}\left[-\left(\frac{1}{2}+m^2\xi_{00}(p,q)\right)+\frac{1}{2}q^2\xi_{10}(p,q)+\frac
	{3}{2}p^2\xi_{01}(p,q)\right],\label{f6}
	\eq
	where $Z_k(p^2,m^2)\equiv\int^1_0 dx\ x^k \ln\left[\frac{m^2-p^2x(1-x)}{m^2}\right]$.
	
	The derivation of the relations (\ref{f1})-(\ref{f6}) can be simply achieved by integration by parts. There is a whole review \cite
	{Orimar2} about these integrals and other integrals with integrands of larger denominators that appear in Feynman diagrams with more external legs.
	
	The result of all finite and regularized divergent integrals from sections \ref{Ss4} is listed below:
	
	\begin{equation}
		\int_k \frac{1}{(k^2-m^2)[(k-p)^2-m^2][(k+q)^2-m^2]}= b \xi_{00}(p,q),
	\end{equation}
	\begin{equation}
		\int_k \frac{k^{\alpha}}{(k^2-m^2)[(k-p)^2-m^2][(k+q)^2-m^2]}= b (p^{\alpha}\xi_{01}(p,q)-q^{\alpha}\xi_{10}(p,q)),
	\end{equation}
	\begin{align}
		\centering
		\int_k \frac{k^{2}}{(k^2-m^2)[(k-p)^2-m^2][(k+q)^2-m^2]}=& I_{log}(m^2)-b Z_0(q^2,m^2)+ b (m^2-p^2)\xi_{00}(p,q)+\nonumber\\
		&+2b(p^2\xi_{01}(p,q)-(p\cdot q) \xi_{10}(p,q)),
	\end{align}
	\begin{align}
		&\int_k \frac{k^{\alpha}k^{\beta}}{(k^2-m^2)[(k-p)^2-m^2][(k+q)^2-m^2]}= \frac{1}{4}g^{\alpha\beta}(I_{log}(m^2)-\upsilon_0)-\frac
		{1}{4}b g^{\alpha\beta} Z_0(q^2,m^2)-\nonumber\\&-b\Big[\frac{1}{2}g^{\alpha\beta}p^2(\xi_{00}(p,q)-3\xi_{01}(p,q)-\xi_{10}(p,q)+2\xi_
		{02}(p,q)+2\xi_{11}(p,q))-\xi_{02}(p,q)p^{\alpha}p^{\beta}+\nonumber\\&+\xi_{11}(p,q)q^{\alpha}p^{\beta}+\xi_{11}(p,q)p^{\alpha}q^{\beta}
		-\xi_{20}(p,q)q^{\alpha}q^{\beta}+(\xi_{10}(p,q)-\xi_{11}(p,q)-\xi_{20}(p,q))g^{\alpha\beta}(p\cdot q)\Big],
	\end{align}
	\begin{align}
		&\int_k \frac{k^{\alpha}k^{2}}{(k^2-m^2)[(k-p)^2-m^2][(k+q)^2-m^2]}=\frac{1}{2}(p^{\alpha}-q^{\alpha})(I_{log}(m^2)-\upsilon_0)+
		\frac{1}{2}b(q^{\alpha}Z_0(q^2,m^2)-\nonumber\\ & -p^{\alpha}Z_0(p^2,m^2))+ b(m^2- q^2)(p^{\alpha}\xi_{01}(p,q)-q^{\alpha}\xi_{10}(p,q))+
		b[q^{\alpha}p^2(\xi_{00}(p,q)-3\xi_{01}(p,q)+\nonumber\\ &-\xi_{10}(p,q)+2\xi_{02}(p,q)+2\xi_{11}(p,q))-2(p\cdot q)p^{\alpha}\xi_{02}(p,q)
		+2 q^2 p^{\alpha}\xi_{11}(p,q)+\nonumber\\ & +2(p\cdot q)q^{\alpha}(\xi_{10}(p,q)-\xi_{20}(p,q))-2 q^2 q^{\alpha}\xi_{20}(p,q))],
	\end{align}
	where  $\int_k\equiv \int^{\Lambda} \frac{d^4 k}{(2\pi)^4}$.

\end{document}